\documentclass{iau-JD}
\usepackage{graphicx}
\def\d{{\rm d}}\def\e{{\rm e}}\def\i{{\rm i}}
\def\kms{\,{\rm km\,s}^{-1}}
\def\kmskpc{\,{\rm km\,s^{-1}\,kpc}^{-1}}
\def\deg{\,{\rm deg}}
\def\pc{\,{\rm pc}}
\def\Myr{\,{\rm Myr}}
\def\spose#1{\hbox to 0pt{#1\hss}}
\def\lta{\mathrel{\spose{\lower 3pt\hbox{$\mathchar"218$}}
     \raise 2.0pt\hbox{$\mathchar"13C$}}}
\def\gta{\mathrel{\spose{\lower 3pt\hbox{$\mathchar"218$}}
     \raise 2.0pt\hbox{$\mathchar"13E$}}}
\def\MNRAS{{\it MNRAS}}
\def\AJ{{\it AJ}}
\def\ApJ{{\it ApJ}}
\def\ApJL{{\it ApJL}}
\def\AaA{{\it A\&A}}
\def\PASJ{{\it PASJ}}
{\newif\ifnotend
\notendtrue
\def\veclist{ABCDEFGHIJKLMNOPQRSTUVWXYZabcdefghijklmnopqrstuvwxyz.}
\def\top#1#2.{#1}
\def\tail#1#2.{#2.}
\loop\expandafter\xdef\csname v\expandafter\top\veclist\endcsname%
{{\noexpand\bf\expandafter\top\veclist}}
\edef\veclist{\expandafter\tail\veclist}
\if\veclist.\notendfalse\fi\ifnotend\repeat}
\let\boldgrk=\gkvecten
\let\boldgrksc=\gkvecseven

\def\gkthing#1{{\mathchoice%
	{\hbox{{\boldgrk\char#1}}}
	{\hbox{{\boldgrk\char#1}}}
	{\hbox{{\boldgrksc\char#1}}}
	{\hbox{{\boldgrksc\char#1}}}}}

\def\vtheta{\gkthing{18}}

\pubyear{2009}
\volume{Volume 15}  %% insert here IAU Highlights of Astronomy Volume Number
\pagerange{1--30}
\date{?? and in revised form ??}
\jname{Highlights of Astronomy, Volume 15}
\editors{Ian F Corbett, ed.}
\begin{document}
 \title[Modelling the Galaxy in the era of Gaia] %% give here short title %%
{Modelling the Galaxy in the era of Gaia\\
Proceedings of Joint Discussion 5
at IAU XXVII, Rio de Janeiro, August 2009}

\maketitle

\begin{abstract}
 The body of photometric and astrometric data on stars in the Galaxy has been
growing very fast in recent years (Hipparcos/Tycho, OGLE-3, 2-Mass, DENIS,
UCAC2, SDSS, RAVE, Pan Starrs, Hermes, ...) and in two years ESA will launch
the Gaia satellite, which will measure astrometric data of unprecedented
precision for a billion stars. On account of our position within the Galaxy
and the complex observational biases that are built into most catalogues,
dynamical models of the Galaxy are a prerequisite full exploitation of these
catalogues. On account of the enormous detail in which we can observe the
Galaxy, models of great sophistication are required. Moreover, in addition to
models we require algorithms for observing them with the same errors and
biases as occur in real observational programs, and statistical algorithms
for determining the extent to which a model is compatible with a given body
of data.

JD5 reviewed the status of our knowledge of the Galaxy, the different
ways in which we could model the Galaxy, and what will be required to extract
our science goals from the data that will be on hand when the Gaia Catalogue
becomes available.

\end{abstract}
\vfill\eject
 \title[Gaia challenge] %% give here short title %%
{The challenge raised by 
Gaia}

\author[A.C. Robin]   %% give here short author list %%
{Annie C. Robin$^1$%
}

\affiliation{$^1$Observatoire de Besan\c{c}on, Institut Utinam, Universit\'e de Franche-Comt\'e, France \break email: annie.robin@obs-besancon.fr\\[\affilskip]
}

\maketitle

\begin{abstract}
 Gaia will perform an unprecedented high quality survey of the Milky Way.
Distances, 3D kinematics, ages and abundances will be obtained, giving access
to the overall mass distribution and to the Galactic potential.  Gaia data
analysis will involve a high level of complexity requiring new and efficient
multivariate data analysis methods, improved modelling of the stellar
populations and dynamical approaches to the interpretation of the data in terms of
the chemical and dynamical evolution of the Galaxy.  
\end{abstract}

The Gaia instruments will perform accurate photometry and astrometry up to
magnitude 20 and spectroscopy to 16th magnitude. The astrometric accuracy is
expected to be at the level of 10, 20, 100$\,\mu$as for stars at $G=10$, 15, 20
resp. (Brown, 2008). This astrometric accuracy will permit  measurement of
parallactic distances up to the Galactic center at the level of 10\%. The
photometry will be accurate at the level of the 0.2$\,$mmag at G=15 and 2.5
millimag at G=20.  From the RVS, the radial velocity errors will be better
than $1\kms$ for bright and cool stars, and at the end of the mission $30\kms$
for stars at $V=16$.  Gaia will also provide astrophysical parameters from the
BP/RP, such as, for a star at $G=15$, effective temperature at the level of 1
to 5\%, gravity to $0.1-0.4\,$dex, metallicity to better than 0.2$\,$dex.
Extinction will be measured on hot stars from RVS with an accuracy of
$0.05-0.1\,$mag. Thus, Gaia provides full characterization for populations to
$G=15$ (accurate distance, age, abundances, 3D velocities). Consequently, ages
will be determined from the astrophysical parameters and stellar evolution
models, as well as relative ages from the elemental abundances. This will
enable us  to trace the chemo-dynamical evolution of different populations. For
example one expects to have about:

\begin{itemize}
 \item 1.5 billion stars ($\sim0.5$\% of the stellar content of the Galaxy)
with photometry, parallaxes and proper motions. Among them, about
$9\times10^8$ stars belonging to the thin disc, $4.3\times10^8$ to the thick
disc, $2.1\times10^7$ to the spheroid and $1.7\times10^8$ to the bulge.

 \item 200 million stars with spectroscopy ($G<16$) (astrophysical parameters
$T_{\rm eff}$, $\log g$, [Fe/H], radial velocity) 

\item 6 million stars with elemental abundances ($G<12$) 

\item Variabilities and binarity.
\end{itemize}

All populations in the Galaxy will be surveyed. Although a limited number of
stars truly in the bulge will be measured with the spectrograph due to
extinction and crowding, (Robin et al, 2005), most stars in the bulge region
will have their parallaxes and proper motions measured.

In the meantime, complementary surveys are planned  that will enhanced the
Gaia outputs, among them LSST, RAVE, PanStarrs and JASMINE. All
these surveys will efficiently complement Gaia, giving better accuracy on
radial velocities for fainter stars, exploring deeper fields, furnishing
denser light curves for variables, revealing dusty regions from the near
infrared, etc. 

Having data sets for billions of objects, covering large numbers of
multi-epoch observables, the data analysis will be a real challenge. The
experience with recent large scale surveys will be of some help but previous
surveys have never reached this level of complexity. The question is: how to
turn Gaia data into a clear understanding of Galactic dynamics and evolution? 

A promising path is to use modelling as a tool for analysis, interpretation
and confrontation between the data and scenarios of formation and evolution
of the Milky Way. Various modelling options are pursued. Since Galactic
evolution leaves traces both in the stellar kinematics and the abundances,
both aspects have to be taken into account. For this reason, the stellar
population-synthesis approach will be valuable, allowing us to compare scenarios
of Galaxy formation and evolution with the Gaia data set by simulating
catalogues with the same observables and comparable accuracies. The Besan\c con
Galaxy model is such a project (Robin et al, 2003), constrained by already
existing surveys like GSC2, DENIS, 2MASS, SDSS, etc., from multi-wavelength
data (from X to infrared) and multivariate catalogues (photometric,
astrometric and spectroscopic). It produces realistic simulations of the
stellar content of the Galaxy with characteristics in agreement with our
present knowledge of Galactic evolution and taking into account
interstellar extinction. It is already used in Gaia preparation and is
planned to be exploited for the data analysis.

Complementary to the synthetic approach, dynamical models allow the
reconstruction of the Galactic potential from the space distribution and the
kinematics of numerous stars or selected tracers. Several dynamical
approaches are being pursued, like the Schwarzshild approach, the torus
method (Dehnen \& Binney 1996), N-body simulations or adaptative N-body
techniques such as the "Made-to-Measure" scheme (Syer \& Tremaine 1996), or
quadratic programming (Dejonghe, 1989). The drawback is that the data sets
are biased with parameters generally not included in the dynamical analysis
-- the first bias being the limiting magnitude but more complex biases also
exist, such as those generated by the fact that the accuracy depends on
magnitudes, colours and position on the sky, and the bias coming from
interstellar extinction. As a consequence, these dynamical modelling
approaches will benefit from being coupled with the synthetic approach.

Using such tools for the interpretation of Gaia data will require  us to develop
efficient methods of multivariate data analysis and model fitting, like
genetic algorithms, or Markov Chain Monte Carlo. Inverse methods will be
attempted but are not easy to handle with the large parameter space furnished by
Gaia.

A huge challenge is then raised by Gaia interpretation. One can expect that
no simplistic model will straightforwardly fit the data. Complemented
by other surveys, the Gaia catalogue will encompass any model view.
Even though imperfect, the modelling will still be useful to i) understand the
imperfections of our knowledge, ii) help to interpret the findings, iii) test
physical scenarios of Galaxy formation and evolution, iv) describe the
dynamics of the system, in relation with the dark matter content, and v)
place the scenario of formation in the cosmological context and constraints
cosmological models.

\vfill\eject
 \title[Dynamics and history of the Milky Way] %% give here short title %%
{Dynamics and history of the Milky Way}

\author[Amina Helmi] %% give here short author list %% 
{Amina Helmi}

\affiliation{Kapteyn Astronomical Institute, University of Groningen, The Netherlands \break email: ahelmi@astro.rug.nl}

\maketitle

\begin{abstract}
The  structure and dynamics of the Galaxy contain
information about both its current workings and its
assembly history. I review our understanding of
the dynamics of the disk and  stellar halo, and sketch how these
may be used to unravel how our Galaxy formed.

\end{abstract}

\firstsection % if your document starts with a section,

\section{Dynamics of the disk(s)}

The structure and kinematics of the Galactic components constrain the
mass distribution and history of the Galaxy.  For example the
vertical dynamics of the thin disk:
\begin{itemize}
\item Puts limits on the distribution of mass in the disk, most of
which is accounted for by the stars (\cite[Kuijken \& Gilmore
1991]{kg}; \cite[Holmberg \& Flynn 2004]{hf}). Their contribution to
the circular velocity (which itself still has large uncertainties,
\cite[McMillan \& Binney 2009]{mcm}) is roughly half of that required
and thus a rounder (dark) component is needed. Also the tilt of the
velocity elliposid rules out very flattened oblate halos
(\cite[Siebert et al. 2008]{siebert}).

\item Its coldness has been used to constrain the amount of recent
merger events (\cite[Toth \& Ostriker 1992]{toth}). Small dark
satellites, which are so abundant in CDM simulations (\cite[Springel et
al.\ 2008]{springel}), do not induce much heating (\cite[Font et
al. 2001]{font}), but mergers of 10-20\% mass ratio can significantly
increase the velocity dispersion of the stars.
\end{itemize}

\cite[Stewart et al. (2008)]{stewart} have found that 70\% of dark-matter
halos similar to that of the Milky Way ($\sim\!10^{12}\,M_{\odot}$) have
experienced a merger with an object of $\sim 10^{11}\,M_{\odot}$ (i.e. mass
comparable to the thin disk's) in the last 10$\,$Gyr. This would therefore be
a plausible origin for the thick disk (\cite[Kazanzidis et al. 2008]{kz};
\cite[Villalobos \& Helmi 2009]{vh}), also if we take into account the age
distribution of its stars (\cite[Bensby \& Feltzing 2009]{bensby}). However, this merger rate does not account for possible
environmental dependences -- the Local Group is in a low density region of the
Universe, which must imply a smaller chance of encounters. Note as well that
this class of mergers are less damaging for gas-rich disks, which is the
relevant case for those lookback times \cite[(Hopkins et al. 2008)]{hopkins}.

Other models, besides the minor-merger scenario, have also been
proposed for the formation of the thick disk. \cite[Abadi et
al. (2003)]{abadi} suggest that it may result purely from the
accretion of satellites on low inclination orbits, while \cite[Brook
et al. (2004)]{brook} find that a thick component might form early on
during gas rich mergers (a different but also gaseous formation
scenario has been put forward by \cite[Bournaud et al.,
2008]{bournaud}). Recently \cite[Sch{\"o}nrich \& Binney (2009)]{sb}
have proposed that the thick disk is composed by stars which have
migrated radially via resonant mechanisms from the inner thin disk.

The dynamics of thick-disk stars encode which of these mechanisms has
been dominant in the formation of this component. Recently,
\cite[Sales et al. (2009)]{sales} have shown that the eccentricity
distribution is a particularly powerful discriminant. In all scenarios
where the majority of the stars are formed in-situ (minor merger, gas
rich mergers or migration), the distribution has a prominent peak at
low eccentricity. On the other hand, when the whole disk is built by
accretion, the eccentricity distribution is predicted to be flatter,
reflecting the range of orbital eccentricities of satellites
found in cosmological simulations.

\section{Dynamics of the stellar halo}\label{sec:trapmode}

The dynamics of halo stars, and especially of those in
streams, can be used to constrain: i) the total mass of the Galaxy
and its spatial distribution (e.g. density and shape of the dark
matter halo); and ii) the merger history of the Galaxy, as
accreted objects will often deposit their debris in this component.

Models of the Sgr streams have yielded conflicting results favouring
spherical, oblate or slightly prolate shapes for the Galactic dark halo
depending on the set of observations used (\cite[Helmi 2004]{helmi04};
\cite[Johnston et al. 2005]{kvj}; however see \cite{law09} who suggest it may
be triaxial). Narrow streams are arguably better-suited to derive the
gravitational potential in the region probed by their orbits (e.g.\ Eyre \&
Binney 2009). \cite{Koposov} have modelled GD-1 and been able to constrain
the circular velocity at the Sun to be $\sim 224 \pm 13\kms$, and the shape
of the potential (including disk and halo) to have a global flattening
$q_\phi \sim 0.87$.

Very high-resolution cosmological CDM simulations in combination with
semi-analytic models of galaxy formation may now be used to make
detailed predictions on the properties of the stellar halo (and in
particular the accreted component, \cite[De Lucia \& Helmi 2008]{gdl};
Cooper et al., in prep.). The most recent such studies show good
agreement with observations, revealing the presence of broad
streams such as those from Sgr (typically originated in massive
recently objects), and very narrow features, akin the Orphan Stream
(\cite[Belokurov et al., 2007]{vasily}). Furthermore,  in these simulations the
very chaotic build up characteristic of the hierarchical structure
formation paradigm endows the stellar halo near the Sun with much 
kinematic substructure (Helmi et al., in prep).

\vfill\eject
 \title[Non-Equilibrium Dynamics] %% give here short title %%
{Non-equilibrium Dynamical Processes in the Galaxy}

\author[Quillen \& Minchev]   %% give here short author list %%
{Alice. C Quillen$^1$ \and Ivan  Minchev$^2$}

\affiliation{
$^1$Dept. of Physics and Astronomy, University of Rochester, \break Rochester, NY 14627, USA \\[\affilskip]
$^2$Universit\'e de Strasbourg, CNRS, Observatoire Astronomique  \break
11 Rue de l'Universit\'e, 67000 Strasbourg, FRANCE
}

\maketitle

\begin{abstract}
Dynamical models have often necessarily assumed that the Galaxy 
is nearly steady state or dynamically relaxed.
However observed structure in the stellar 
metallicity, spatial and velocity distributions
imply that heating, mixing and radial migration
has taken place.  
%GAIA observations will reveal even more 
%structure in these distributions. 
Better comprehension of non-equilibrium processes
will allow us to not only better understand
the current structure of the galaxy but its past evolution.
%I review the dynamics of processes
%such as related to spiral arm growth, evolution and decay, 
%bar growth and evolution, interplay between
%these structures, mergers and nearby
%satellite galaxies or subhalos and discuss how they
%might leave signatures in the stellar distributions.
%\keywords{Keyword1, keyword2, keyword3, etc.}
\end{abstract}


%\section{Introduction}

During a Hubble time the Milky Way disk at the radius of the Sun
has only had time to rotate 40 or 50 times.  
There is little time for dynamical relaxation. 
As larger and more precise surveys are conducted we expect
even more structure to be revealed in the stellar abundance and phase
space distributions. 
As there is little time for relaxation, structure in the
phase space distribution depends on the evolution 
of the Galaxy.
%
%Here I mention some non-equilibrium processes and how
%they might affect the structure of the Galactic disk.

Resonances with a bar or spiral structure cause stars
to move in non-circular orbits.  
Libration times in Lindblad resonances can be long so evolution could 
take place in the non-adiabatic limit.   
\cite{minchev09b} proposed that
the division between the Pleiades and moving groups
in the solar neighborhood 
is associated with librations in the 2:1 Lindblad
resonance with the Galactic bar (see Figure 1). 
These oscillations are also seen
as long lived R1/R2 asymmetric ring structures in test particle
simulations of bar growth (Bagley et al.~2009).

If pattern speeds vary, then either particles are trapped into
resonance or heated as they cross the resonance.
When particles are trapped into resonance their eccentricity depends
on the total pattern speed change after capture. 
When particles cross the resonance their eccentricity
can be predicted from the resonance strength and order.
\cite{bagley09} suggested that the morphology
of ring structures associated with a bar 
depends on bar pattern speed variations
since growth. Peanut shaped bulges can 
also be modeled with a resonant trapping model \cite{quillen02}.
When there is more than one perturbation, chaotic heating
occurs in resonances (\cite{quillen03,minchev06}).
%Past growth and decay of spiral structures can 
%leave resonant imprints in the velocity distribution (\cite{desimone}).
%Resonances can also be responsible for radial migration (\cite{sellwood}).

Since resonances are often narrow, they can be used to
place tight constraints on their pattern speed. 
Their location in the galaxy could be used to measure the
pattern speed of a distant spiral pattern with a deep radial velocity 
survey \cite{minchev08}.
Resonances occur when the sum of integer multiple of a star's
orbital frequencies is equal to an integer multiple of
a perturbing frequency such as a planet's mean motion
or the pattern speed of a bar.  
%In the asteroid belt resonances depend on the semi-major axis as that
%sets the orbital period. 
In the Galaxy the orbital period
is estimated from the tangential or $v$ velocity component at a particular
location.  Thus resonances can be located on a $u,v$ plane velocity
distribution.
Divisions between streams 
can be used to estimate bar or spiral pattern speeds
(\cite{dehnen99,quillen05}).
By matching both the velocity distribution in the solar neighborhood
and simulated Oort function measurements (that depend on velocity
gradients) \cite{minchev07a} placed an even tighter constraint
on the bar pattern speed.

Differences between orbital frequencies cause the velocity and
spatial distribution of stars in a narrow region in phase space 
to spread.  This process is 
called ``phase wrapping'' and can be used to estimate 
the time since a merger occurred \cite{gomez09}.  Uneven distributions
in phase space could also be caused by large scale perturbations
to the disk. The timescale for them to wrap 
places constraints on the time since 
perturbation (\cite{minchev09}).
These scenarios are proposed explanations for high
velocity streams in the thick disk of the Galaxy.
Both mergers (\cite{quillen09}) and resonances (\cite{sellwood}) cause
radial migration. Future work can better explore 
the relation between structure in the phase space and abundance
distributions.

%\section{Discussion}

%In summary the dynamics of the disk is rich.  
In summary, dynamical structures and events leave signatures in the stellar 
distributions.  Precise measurements can be made as observations
and associated models become more comprehensive.
Unveiling the current and past structure and evolution of
the Galaxy will be increasingly exciting in the coming decade.

\begin{figure}
\includegraphics[scale=0.39]{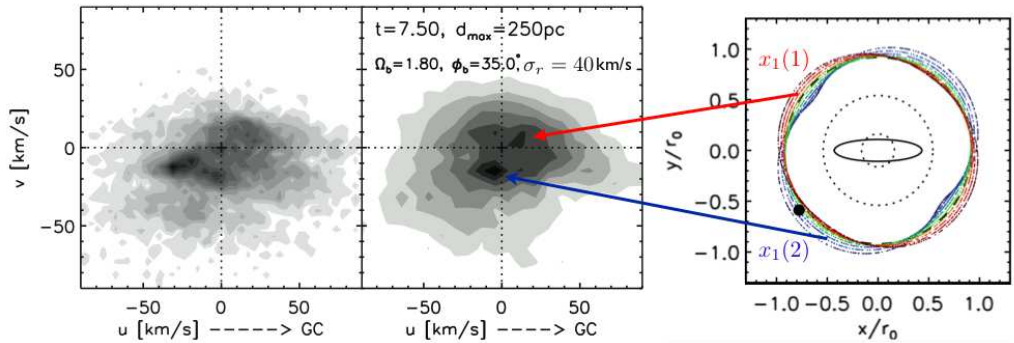}
\caption{Orbits associated with the Bar.
On the left we show the Hipparcos stellar 
velocity distribution (\cite{dehnen98}).
The middle panel shows a model distribution by \cite{minchev09b}.
On the right we show orbits associated with the 
Sirius and Pleiades moving groups.
}
\label{fig:fig1}
\end{figure}

%\subsection{Basic properties}

%\begin{acknowledgments}
%We would like to acknowledge the useful comments of a referee.
%\end{acknowledgments}

{}
\vfill\eject
 \title[Star-formation histories, metallicity distributions and luminosity functions] %% give here short title %%
{Star-formation histories, metallicity distributions and luminosity functions}

\author[Rosemary F.G.~Wyse]   %% give here short author list %%
{Rosemary F.G.~Wyse}

\affiliation{Department of Physics \& Astronomy, Johns Hopkins University, Baltimore, MD 21218, USA 
\break email: wyse\@pha.jhu.edu}

\maketitle

\begin{abstract}
A selection of topics was discussed, but 
given the limits of space, I discuss here only the IMF in any depth,  with only a brief supplementary comment on the bulge. 
\end{abstract}

\firstsection
\section{The Stellar Initial Mass Function}

The stellar Initial Mass Function is a fundamental aspect of star
formation. Variations in the IMF are predicted in many theories,
particularly those that invoke the Jeans mass, and associated
dependences on cooling rates and pressure. The environment of the
Galactic Center is rather different from the solar neighbourhood, and
the dense, massive, young star clusters there provide an interesting
test of the variation, or not, of the massive-star IMF. Indeed, early
star-count observations had suggested a flatter IMF in the Arches star cluster,
some $25\pc$ in projection from the Galactic Center and with age $\lta3\Myr$, than in the solar
neighbourhood (Figer et al.~1999). However, the most recent data,
obtained with the NACO Adaptive Optics imager on the VLT, are entirely
consistent with a standard, Salpeter (1955), slope to the IMF above
$10\,M_\odot$ (Espinoza, Selman \& Melnick 2009), {\it modulo\/} a level
of mass segregation in the inner regions.

The level of variation in generations of massive stars that have since
exploded as core-collapse supernovae can be constrained through
analysis of the elemental abundances in low-mass stars that they
pre-enriched. In general, if the massive-star IMF is biased towards
more massive stars, the predicted ratio of alpha-elements to iron will
be higher, due to the dependence of the nucleosynthetic yields on
progenitor mass, within the range of core-collapse progenitors, $\sim
10\,M_\odot$ to $\sim 50\,M_\odot$ (e.g.~Kobayashi et
al.~2006). Stars formed early in the history of self-enrichment of a
system will show signatures of only core-collapse supernovae, and when
these stars can be identified, they show remarkably little scatter in
most elemental abundance ratios, even down to very low levels of
[Fe/H] (e.g.~Cayrel et al. 2004).  This constancy, within estimated
errors, implies not only an invariant massive-star IMF, but also
efficient mixing so that IMF-averaged yields are achieved (e.g.~Wyse
\& Gilmore 1992; Fran\c{c}ois et al.~2004).  It should be noted that the conclusion of the invariance
of the IMF from the lack of scatter is not dependent on the details of the supernova models, nor on  a
chemical evolution model, but is a robust conclusion.  Of course,
determination of what the IMF is from the value of the
elemental ratios would be sensitive to models.

These
conclusions hold for stars in diverse environments such as the old,
metal-rich bulge (e.g.~Fulbright, McWilliam \& Rich 2007) and
metal-poor halo. Even in the metal-poor dwarf spheroidals, where the
bulk of the member stars show lower values of [$\alpha$/Fe] (e.g.~Venn
et al.~2004), consistent with their broad internal spread of ages and likely
self-enrichment by Type Ia supernovae (see Unavane, Wyse \& Gilmore 1996),
the most metal-poor -- and presumably oldest -- stars show the same
enhanced ratio as in the field halo (e.g.~Koch et al.~2008). 

Extending this analysis to larger samples of extremely metal-poor
stars is possible through exploitation of the database from the RAVE
moderate-resolution spectroscopic survey of bright stars ($I < 12$;
see Steinmetz et al.~2006 for an introduction to the survey), for
which 4--8m-class telescopes suffice for follow-up high-resolution
data.  Preliminary results (Fulbright, Wyse, Grebel et al., in prep)
are very encouraging.  Extremely metal poor stars, with [Fe/H] $ <
-3\,$dex, have also been identified in dwarf spheroidal galaxies
(e.g. Kirby et al.~2008; Norris et al.~2008), predominantly in the
`ultra-faint' systems discovered by analysis of the imaging data of
the Sloan Digital Sky Survey (e.g.~Belokurov et al.~2006, 2007).

As noted below, the
ages of the bulge and halo stars are equal to look-back times
corresponding to redshifts of 2 and above, implying constancy of the
massive star IMF back to these early stages of the Universe, in a wide range of physical conditions.

The low-mass luminosity function can be determined in most systems by
straightforward star counts, and again the constancy or otherwise of the IMF
can be tested in systems that probe a wide range of physical parameters: the
high-metallicity, old, dense stellar bulge; the low-metallicity, old,
diffuse, dark-matter dominated dwarf spheroidals; the low-metallicity, old,
dense Globular star clusters; and lastly the young disk star and open
clusters.  Again, all is consistent with an invariant low-mass IMF (e.g.~Wyse
2005). 

In terms of modelling the Galaxy, the IMF should be held fixed. 

%\cite[Callan \etal\ (1991)]{Callan91} 

\section{The Bulge}

The conclusion that the bulge consists of exclusively old stars -- at
least those regions probed by low-reddening windows -- is not new, but
has been given considerably more weight by a recent HST-based analysis
by Clarkson et al.~(2008) in the Sgr low-extinction window.  These
authors used multi-epoch observations to identify foreground disk
stars, through their proper motions.  After removal of these younger
stars, the colour-magnitude diagram remaining is that of an old,
metal-rich population.  However, much more data are required to map the
stellar population in the bulge, and understand how it connects to the
stellar halo, and to the inner thin and thick disks.

\vfill\eject
 \title[Interstellar medium] %% give here short title %%
{Structure and evolution of the Milky Way: the interstellar medium perspective}

\author[F. Boulanger]{Fran\c cois Boulanger}
\affiliation{Institut d'Astrophysique Spatiale\break email:
  francois.boulanger@ias.u-psud.fr}

\maketitle

\begin{abstract}
The Herschel and Planck satellites have started imaging the 
sky at far-IR to mm wavelengths 
with an unprecedented combination of sky and spectral coverage, angular resolution,
and sensitivity, thus opening the last window of the
electromagnetic spectrum on the Galaxy. 
Dedicated observing programs on Herschel and the Planck all-sky 
survey will  provide the first complete view at cold dust across the 
Galaxy, opening new perspectives on the structure and dynamical 
evolution of the Milky Way relevant to Gaia. 
The analysis and modelling of these observations will contribute to 
our understanding of two key questions:
how do stars form from interstellar matter? 
how are the interstellar medium and the magnetic field dynamically 
coupled? The comparison with Gaia observations will contribute to build 
a 3D model of the Galactic extinction taking into account dust 
evolution  between ISM components 
\end{abstract}

\firstsection % if your document starts with a section,

\section{Introduction}

The Gaia mission will produce an extraordinary stellar database and we have
to consider how the data may be used to build a dynamical model of the Milky
Way that will unravel its past history.  To understand the evolution of the
Galaxy, we need a model that describes physically how star formation proceeds
from the chemical and thermo-dynamical evolution of interstellar matter.
This is a vast field of research. At the meeting I highlighted the prospect
of imminent advances made possible by the successful launch of the Herschel
and Planck satellites. Here I focus on two main topics related to star
formation: the inventory of the cold interstellar medium across the galaxy,
and the structure of the Galactic magnetic field and its coupling to
interstellar matter.

\section{Cold interstellar matter across the Galaxy}\label{sec:greenfun}

Herschel and Planck surveys will provide the small-scale (down to the
detection of individual pre-stellar cores) and global views of the
distribution of cold interstellar matter in the Galaxy. Both missions will
image the far-IR emission from large ($> 10\,$nm) grains that account for the
bulk of the dust mass. The spectral coverage will allow us to determine the
dust temperature and thereby infer dust column densities and masses. With an
empirical determination of the dust-to-gas mass ratio, the infrared
brightness becomes a tracer of interstellar gas. It complements usual
interstellar matter tracers, like the H$\,$I and CO line emission, in a unique
way because the dust emission is independent of the chemical composition and
physical conditions of the gas. For the first time we will have access to a
complete inventory of cold interstellar matter in the Galaxy. This step
forward opens several key perspectives.

The dust temperature may be used to identify the cold infrared emission from
dense condensations within molecular clouds. In doing this, we will quantify
the mass and distribution of matter that is presently susceptible to collapse
into stars.  The data will be sensitive enough to look for dust emission from
High Velocity Clouds, the Magellanic Stream and the outer disk. This will
help us to quantify the mass-inflow rate available to sustain star formation
over the past history of the Galaxy.  The dust seen in emission by Herschel
and Planck is the dust that makes the extinction at optical and near-IR
wavelengths. The data analysis will involve the characterisation of dust
evolutionary processes that may account for the observed variations in the
optical and near-IR dust extinction curve.  The outcome of these dust studies
will need to be taken into account to build the 3D model of extinction for
Gaia. 

\section{The Galactic magnetic field}\label{sec:magnetic}

The Galactic magnetic field and cosmic-rays are tied to the interstellar gas.
Their dynamical coupling is a prime facet of interstellar-medium physics, and
many questions remain quantitatively open due to the paucity of data on the
small-scale structure of the magnetic field.  Planck will map the
polarisation of the dust and of the synchrotron emission. The two emissions
provide complementary perspectives on the structure of the Galactic magnetic
field. Dust grains, unlike relativistic electrons, are coupled to the
interstellar gas. Thus, the dust polarisation traces the magnetic field
within the thin Galactic disk where matter is concentrated and within
interstellar clouds, while the synchrotron polarisation probes the field over
the whole volume of the Galaxy up to its halo. The novelty of upcoming
observations ensures major progress in our understanding of the magnetised
Galactic interstellar medium.

Polarisation of the dust emission results from the presence in the ISM of
elongated grains with a preferred orientation.  Several alignment mechanisms
have been proposed. They are expected to work through interstellar clouds
even if their efficiency may depend on the gas density and radiation field.
It is thought to be generally true that the magnetic field acts on elongated
grains so they spin with their long axes perpendicular to the field. In
this case, the direction of polarisation in emission is perpendicular to
$B_{\bot}$ the magnetic field component in the plane of sky.  The degree of
polarisation depends on dust properties (e.g. which grains are aligned), the
efficiency of the alignment mechanism and the structure of the magnetic field
within the beam. 

Measurement by Planck of the polarisation of the thermal emission from
aligned grains provide an unprecedented means to map continuously the
orientation of the magnetic field within the ISM, from diffuse clouds to
dense molecular gas.  The Galactic magnetic field is commonly described as a
vector sum of a regular and a random component.  A first goal will be to
complement existing models of the regular component.  To fully describe the
ordered field, we will face two open questions:  (i) what is the impact of nearby
bubbles powered by massive star associations on the field structure?  (ii) how is
the field within the thin Galactic disk, where gas and star formation is
concentrated, connected to the thicker disk and the Galactic halo?  The
turbulent component results from the dynamical interaction between the field
and interstellar turbulence.  The data will also allow us to study the
geometry of the magnetic field in relation to the density structure and
kinematics of interstellar clouds derived from dust and gas maps.  The degree
of randomness in the magnetic field orientation may be combined with Doppler
measurements of the turbulent gas velocity to measure the magnetic field
intensity and quantify the dependence on gas density.  These investigations
will test theoretical and numerical studies, stressing the importance of the
magnetic field in the dynamical evolution of the interstellar medium and the
regulation of the star-formation efficiency. 

 \vfill\eject
 \title[New Frontiers in Galactic Archaeology] %% give here short title %%
{The Milky Way Halo and the First Stars:  New Frontiers in Galactic Archaeology}

\author[Beers et al.]   %% give here short author list %%
{Timothy C. Beers$^{1,2}$, Jason Tumlinson$^3$, Brian O'Shea$^{1,2}$,
Carolyn Peruta$^{1,2}$, Daniela Carollo$^{4,5}$} 

\affiliation{$^1$Department of Physics \& Astronomy, Michigan State
University, \break  email: beers@pa.msu.edu \\[\affilskip]
$^2$Joint Institute for Nuclear Astrophysics\\[\affilskip]
$^3$Space Telescope Science Institute\\[\affilskip]
$^4$Research School of Astronomy \& Astrophysics, ANU, Australia\\[\affilskip]
$^5$INAF, Osservatorio Astronomico di Torino}

\maketitle

\begin{abstract}

We discuss plans for a new joint effort between observers and
theorists to understand the formation of the Milky Way halo back to
the first epochs of chemical evolution. New models based on
high-resolution N-body simulations coupled to simple models of
Galactic chemical evolution show that surviving stars from the epoch
of the first galaxies remain in the Milky Way today and should bear
the nucleosynthetic imprint of the first stars. We investigate the key
physical influences on the formation of stars in the first galaxies
and how they appear today, including the relationship between cosmic
reionization and surviving Milky Way stars. These models also provide
a physically motivated picture of the formation of the Milky Way's
``outer halo," which has been identified from recent large samples of
stars from SDSS. The next steps are to use these models to guide
rigorous gas simulations of Milky Way formation, including its disk,
and to gradually build up the fully detailed theoretical ``Virtual
Galaxy" that is demanded by the coming generation of massive Galactic
stellar surveys.

\end{abstract}

%\section{Summary}

The explosion of detailed astrometric, photometric, and spectroscopic
data for stars in the Milky Way (and Local Group galaxies) that is
coming upon us both now (e.g., SDSS/SEGUE, RAVE) , and in the near
future (PanSTARRS, SkyMapper, Gaia, LSST, SIM Lite) will fundamentally
change our vision of galaxy formation and evolution. Full exploitation 
of this wealth of new information requires 
the development of sophisticated numerical models capable of producing
testable predictions against which the observations can be compared.
We have initiated one such effort, foreshadowed by the work of
\cite[Tumlinson (2006)]{Tum2006}, and continued by \cite[Tumlinson (2009)]{Tum2009}. The new 
models predict the locations, kinematics, and chemistry of the stars
in the Galactic halos, with different assumptions concerning the
nature of the first stars and their effects on subsequent stellar
generations. We are also exploring new methods for visualizing both
the predictions and the existing databases. The first efforts will
compare expectations with observed differences in the inner- and
outer-halo populations, e.g., as in \cite[Carollo et al. (2007)]{Carollo2007}.

{}
\vfill\eject
 \title[Structure of the Galactic disc(s)] %% give here short title %%
{Physics and Structure of the Galactic disc(s)}

\author[Ralph A. Sch{\"o}nrich]{Ralph A. Sch{\"o}nrich}

\affiliation{Max Planck Institute for Astrophysics, \\ Garching,
D-85741, Garching, Germany \\ email: {\tt rasch@mpa-garching.mpg.de}}

\maketitle

\begin{abstract}
The model of Sch{\"o}nrich \& Binney (2009) offers new ways to understand the
chemo-kinematic structure of the solar neighbourhood in the light of radial
mixing. The combination of chemical information with rich kinematic data
reveals a still hardly explored abundance of interconnections and structures
from which we can learn about both the physics and history of our Galaxy.
Large upcoming datasets can be used to
improve estimates of central parameters, to shed light on the Galaxy's
history and to explore the unexpected way of understanding the well-known
division of the Galactic disc yielded by the new model. \keywords{
%%galaxies: structure, galaxies: evolution, galaxies: abundances, galaxies: kinematics and dynamics, 
Galaxy: structure, Galaxy: evolution, Galaxy: abundances, Galaxy: kinematics and dynamics} 
\end{abstract}

Radial mixing has been a vastly neglected process in modelling galactic discs
until it was shown to be crucial for disc evolution from a theoretical point
of view by \cite{SeB02}. \cite{SB09a} demonstrated that under regular assumptions
about star formation and disc structure, a perfect fit to the metallicity
distribution of the Geneva-Copenhagen Survey (\cite[Nordstr{\"o}m et
  al. 2004]{Nord04}, \cite[Holmberg et al. 2007]{HolmbergNA}) was possible if
radial mixing was allowed for, and that no acceptable fit could be achieved
without it. This new model of chemical evolution gives a completely different
history of the disc compared to classical approaches. Since a large range of
different galactocentric radii contribute their populations to local
datasets, there is no need for the local star-forming interstellar medium to
have followed the observed density ridges in the [$\alpha$/Fe], [Fe/H] plane.
Among the most important differences between classical modelling without
radial mixing and models including radial migration are the predicted
correlations between chemistry and kinematics along the thin-disc ridge line.
In larger datasets containing [$\alpha$/Fe] ratios, such as that of
\cite{BoMa05}, thin-disc stars can be picked out from stars belonging to the
thick disc and halo by selecting for low [$\alpha$/Fe] and dropping obvious
halo stars. In this subsample one finds a highly significant increase of
rotational velocity along the thin-disc ridge line moving from higher to
lower metallicities, in nice concordance with the SB09 model.  In the
classical models the metal-poor stars should be the oldest thin-disc stars,
so the increase in rotational velocity should be associated with an increase
in velocity dispersion. The data do not show a significant increase towards
lower metallicities, so the classical view is in conflict with the data.

 \vfill\eject
 \title[Gas flows within the Galaxy] %% give here short title %%
{Gas flows within the Galaxy}

\author[Combes]   %% give here short author list %%
{Francoise Combes}

\affiliation{LERMA, Observatoire de Paris,
61 Av. de l'Observatoire, F-75014, Paris \break email: francoise.combes@obspm.fr}

\maketitle

\begin{abstract}
 In the recent years, more and more sophisticated models have been proposed for the gas
distribution and kinematics in the Milky Way, taking into account the main bar, but
also the possible nuclear bar, with the same or different pattern-speed.
I  review the success and problems encountered by the models, in particular in
view of the new discovery of a symmetrical far-side counterpart of the 3$\,$kpc arm.
 The inner part, dominated by the bar, and the outer parts, dominated
by the spiral arms, can be observed from a virtual solar position,
and the errors coming from kinematical distances are evaluated.
The appearance of four arms could be due to a deprojection bias.
\end{abstract}

\firstsection % if your document starts with a section,
\section{Introduction}

Reconstruction of the Galaxy's
spiral arms is difficult given our internal perspective. Distances of the various features or
objects are derived through a kinematical model, with near-far ambiguities
inside the solar circle. One of the most successful models is that from
Georgelin \& Georgelin (1976) of four tightly-wound arms, traced by OB
associations, optical or radio HII regions and molecular clouds. The best
tracer of the Galaxy structure is the gas, atomic (HI, Liszt \& Burton 1980)
and molecular (CO surveys, Dame et al 2001), because of its low velocity
dispersion, and its confinement to the plane.

Position-velocity (P-V) diagrams are particularly instructive, revealing the
high-velocity ($\Delta V=560\kms$) Central Molecular Zone (CMZ) near zero
longitude, with a molecular ring, connecting arm, 3$\,$kpc arm, etc.
 The existence of a bar has long been suspected from non-circular motions
towards the center, and has been directly confirmed by COBE and 2MASS (e.g.
Lopez-Corredoira et al 2005). Near-infrared images show clearly the peanut
bulge, which is thought to be formed through vertical resonance with the bar
(e.g. Combes et al 1990).  The CMZ has a peculiar parallelogram shape in the
P-V diagram (Bally et al 1988), that has been first interpreted in terms of
cusped $x_1$ and almost circular $x_2$ periodic orbits, and associated gas
flows, by Binney et al (1991).
 Then Fux (1999) carried out fully self-consistent N-body and hydrodynamical simulations of
 stars and gas to form a barred spiral, and fit the Milky Way. He succeeded
remarkably to reproduce the HI and CO P-V diagrams with a bar pattern speed
of about $40\kmskpc$, implying corotation at 5$\,$kpc, and an ILR producing
the $x_2$ orbit inside 1$\,$kpc. The spiral structure has essentially
2 arms starting at the end of the bar.

\section{More recent developments}
 Both external galaxies and simulations frequently show evidence for several
pattern speeds in the disk and a spiral is expected to rotate slower than the
bar (Sellwood \& Sparke 1988). For example, by modelling gas flow in a fixed
potential Bissantz et al (2003) conclude that in the Galaxy $\Omega_{\rm p}=
60\kmskpc$ and $20\kmskpc$ for the bar and spiral, respectively.  Amores et
al (2009) notice that there is a ring gap in the HI distribution at about
8.3$\,$kpc, outside the solar circle at 7.5$\,$kpc, and propose that it
corresponds to the corotation (CR) of the spiral, which will then have a
pattern speed of $\sim25\kmskpc$. Simulations of gas in a barred spiral do
not show ring gaps at CR, but depopulated regions at the corresponding
Lagrangian points, which could correspond to the observations; there is no
gap in the azimuthally averaged HI and CO gas surface density.

In the 2MASS star counts, a nuclear bar has been found by Alard (2001), and
there is a corresponding CO nuclear bar (Sawada et al 2004). New simulations
of gas flow in a two-bar models have been done by Rodriguez-Fernandez \&
Combes (2008), who find a best fit when the two bars are nearly
perpendicular, and the bar-spiral pattern is about $35\kmskpc$ (similar to
Fux, 1999).  The model shows the far-side twin of the 3$\,$kpc arm, which
has just been discovered in the CO P-V diagram (Dame \& Thaddeus 2008).  It
reproduces also the connecting arm (characteristic leading dust lanes along
the bar).  No evidence is found of lopsidedness in the stellar potential, and
the CO lopsidedness must be a purely gaseous phenomenon.

\section{Remaining problems}
 New reconstruction of the spiral structure in the galactic plane have been
attempted in the HI gas (Levine et al 2006), and in the CO gas (Nakanishi \&
Sofue 2006, Englmaier et al 2009).  The best fit could be two arms, starting
at the end of the bar, with a pitch angle of 12$^\circ$, although four arms
are still possible. Pohl et al (2008) have tried novel deprojections, by
simulating the gas flow with SPH in a bar potential, and obtaining distances
with a kinematic model derived from the non-circular velocity field obtained.
A test of the procedure with a 2 arms+bar fiducial model, with only one
pattern speed, retrieves after deprojection a four arms spiral.  These recent
efforts demonstrate further the difficulty of disentangling distances and
dynamical effects. It is still possible that several patterns exist in the
Galaxy.  Other prominent features have not yet been interpreted, such as the
warp or tilt of the nuclear structure, or its lopsidedness.

\vfill\eject
 \title[Mapping the Milky Way with SDSS, Gaia and LSST] %% give here short title %%
{Mapping the Milky Way with SDSS, Gaia and LSST}

\author[\v{Z}eljko Ivezi\'{c}$^1$ (for the LSST Collaboration)]   %% give here short author list %%
{\v{Z}eljko Ivezi\'{c}$^1$ (for the LSST Collaboration)}

\affiliation{$^1$Department of Astronomy, University of Washington,
Seattle, WA 98155, USA \break email: ivezic@astro.washington.edu}

\maketitle

\begin{abstract}
We summarize recent work on the Milky Way ``tomography'' with SDSS
and use these results to illustrate what further breakthroughs can be
expect from Gaia and the Large Synoptic Survey Telescope (LSST). 
LSST is the most ambitious ground-based survey currently planned in 
the visible band. Mapping of the Milky Way is one of the 
four main science and design drivers. The main $20\,000\deg^2$. 
survey area will be imaged about 1000 times in six 
bands ($ugrizy$) during the anticipated 10 years of operations,
with the first light expected in 2015. Due to Gaia's superb 
astrometric and photometric accuracy, and LSST's significantly 
deeper data, the two surveys are highly complementary: Gaia 
will map the Milky Way's disk with unprecedented detail, 
and LSST will extend this map all the way to the halo's edge.
\end{abstract}

\firstsection

\section{The Milky Way Tomography with SDSS} 

With the SDSS data set, we are offered for the first time an opportunity to examine in situ
the thin/thick disk and disk/halo boundaries over a large solid angle, using millions of stars.
In a three-paper series, Juri\'{c} et al. (2008), Ivezi\'{c} et al. (2008a) and Bond et al. (2009)
have employed a set of photometric parallax relations, enabled by accurate SDSS multi-color 
measurements, to estimate the distances to tens of millions of main-sequence stars. 
Photometric metallicity estimates based on the $u-g$ colors are also available for about
six million F/G stars, and proper motions based on a comparison of SDSS and the Palomar
Observatory Sky Survey positions are available for about 20 million stars.

With these distances, accurate to $\sim$10\%, the stellar distribution in the multi-dimensional 
phase space can be mapped and analyzed without any additional assumptions. The adopted
analytic models and a computer code ({\it galfast}\footnote{See http://hybrid.mwscience.net.}) 
that summarize these results, can be used to generate mock catalogs for arbitrary depths and
photometric systems (including kinematic quantities). They also enable searches for substructure
by subtracting the smooth background distributions. Indeed, a lot of substructure is
seen in the data in all projections of the parameter space (spatial distributions, kinematics,
metallicity distribution). 

The extension of observations for numerous main-sequence stars to distances
up to $\sim10\,$kpc represents a significant observational advance, and
delivers powerful new constraints on the dynamical structure of the Galaxy.
For example, most stars observed by the Hipparcos survey are within $\sim$100
pc (Dehnen \& Binney 1998).  In less than two decades, the observational
material for such in situ mapping with main-sequence stars has progressed
from first pioneering studies based on only a few hundred objects (Majewski
1993), to over a thousand objects (Chiba \& Beers 2000), to the massive SDSS
data set.

These new quantitative results enable fairly robust predictions for the performance of 
new surveys, such as Gaia and LSST (Eyer et al., in prep.). Due to Gaia's superb
astrometric and photometric measurements, and LSST's significantly 
deeper data, the two surveys will be highly complementary: Gaia will map 
the spatial, metallicity and kinematic distributions of stars in
the Milky Way's disk with unprecedented detail, and LSST will extend 
these maps all the way to the halo's edge, and will obtain large local
samples of intrinsically faint sources such as L, T and white dwarfs. 
We briefly describe LSST in the next section. 

\section{Brief Overview of LSST}

LSST will be a large, wide-field ground-based system designed to 
obtain multiple images covering the sky that is visible from
Cerro Pach\'{o}n in Northern Chile. The LSST design is driven by 
four main science themes: constraining dark energy and dark matter, 
taking an inventory of the Solar System, exploring the transient 
optical sky, and mapping the Milky Way. The current baseline 
design, which envisages  an 8.4$\,$m (6.7$\,$m effective) primary mirror, a 9.6 deg$^2$ 
field of view, and a 3,200 Megapixel camera, will allow about 
10,000 square degrees of sky to be covered using pairs of 15-second 
exposures in two photometric bands every three nights on average. 
The system is designed to yield high image quality as well as superb 
astrometric and photometric accuracy. The survey area will include 
30,000$\,$deg$^2$ with $\delta<+34.5^\circ$, and will be imaged multiple 
times in six bands, $ugrizy$, covering the wavelength range 320--1050
nm. About 90\% of the observing time will be devoted to a deep-wide-fast 
survey mode which will observe a $20\,000\deg^2$ region about 1000 times 
in the six bands during the anticipated 10 years of operations. These 
data will result in databases including 10 billion galaxies and a 
similar number of stars, and will serve the majority of science programs. 
The remaining 10\% of the observing time will be allocated to special
programs such as Very Deep and Very Fast time domain surveys.
More information about LSST can be obtained from www.lsst.org and
Ivezi\'{c} et al. (2008b). 

Each 30-sec observation will be about 2 mag deeper than SDSS imaging, 
and the repeated observations will enable proper-motion and
trigonometric parallax measurements to $r=24.5$, about 4-5 mag fainter 
limit than to be delivered by Gaia, and the coadded LSST map will reach 
$r=27.5$. Due to Gaia's superb astrometric and photometric accuracy, 
and LSST's significantly deeper data, the two surveys will be highly 
complementary.  As shown by Eyer et al., in the range $19<r<20$
Gaia's and LSST errors are fairly similar (within a factor of $\sim2$). 
Towards brighter magnitudes, Gaia's error significantly decrease
(by about a factor of 10 already at $r\sim14$), and towards fainter
magnitudes, LSST will smoothly extend the Gaia's error vs. magnitude 
curves by over 4 mag.

 \vfill\eject
 \title[Schwarzschild Models]{Schwarzschild Models for the Galaxy}

\author[Chaname]{Julio Chanam\'e}

\affiliation{Carnegie Institution of Washington, Department of Terrestrial Magnetism \break 5241 Broad Branch Rd., Washington DC 20015, USA \break email: jchaname@dtm.ciw.edu}

\maketitle
\begin{abstract}

  Schwarzschild's orbit-superposition technique is the most
  developed and well-tested method available for constraining the
  detailed mass distributions of equilibrium stellar systems.  Here I
  provide a very short overview of the method and its existing
  implementations, and briefly discuss their viability as a tool for
  modeling the Galaxy using Gaia data.

%\keywords{Keyword1, keyword2, keyword3, etc.}
\end{abstract}

\firstsection % if your document starts with a section,
\section{Introduction}

Models are used to relate observations to theoretical constructs such as the
phase-space distribution function. A number of simplifying assumptions are
required to make the modeling process tractable and these assumptions can
have profound implications for the inferred mass distribution.  A delicate
balance must be struck between the available observations and the complexity
of the models fitted to them.

The Gaia mission will require modeling techniques that are capable of
handling huge numbers of measurements, while taking full advantage of the
high precision of the data.  Existing implementations of the
orbit-superposition method already fulfil some of these requirements but do
have limitations.  For example, while an assumption regarding the
geometry of the gravitational potential is inescapable, Schwarzschild models
can be built that are completely free from assumptions regarding the detailed
orbital structure, which generally have the largest impact on the derived
mass distribution.

\subsection{Schwarzschild's technique}

Given an orbit library for an assumed gravitational potential, the
orbit-superposition technique \cite[(Schwarzschild 1979)]{sch79} finds the linear sum of those
orbits that best reproduces the available observations.  The success of the
method relies on two aspects: (1) that the stellar system can be safely
considered to be in equilibrium, and (2) that the orbit library is
sufficiently comprehensive.  If these two conditions are satisfied, the
method is very general and free from most assumptions.  Even the required
assumption of a given geometry for the gravitational potential is in practice removed by the iterative
nature of the technique, which calls for the construction of Schwarzschild
models for an entire grid of potentials, with the final model the one that
best fits the data.  Of course a Schwarzschild model only provides a snapshot
of the current dynamical state of the system, and the question of stability
must be addressed by other means.

Schwarzschild models have been successfully used to constrain the dark-matter
halos of galaxies (e.g., Rix et al. 1997, Thomas et al. 2005), to weigh
supermassive black holes at the centers of both galaxies (e.g., van der Marel
et al. 1998, Gebhardt et al. 2000, 2001) and globular clusters (Gebhardt,
Rich, \& Ho 2005). They have also been used to study the dynamics of star
clusters (e.g., van de Ven et al. 2006).  More relevant to the subject at
hand, orbit-superposition models have also been used to study the dynamics of
the Galactic bulge (Zhao 1996, H$\ddot{\rm a}$fner et al. 2000).  Existing
implementations of the Schwarzschild method are usually classified/labeled
according to the geometry of the stellar systems they can be applied to
(spherical, axisymmetric, triaxial), and to the type of dataset they are
designed to handle (continuous or discrete; see Chanam\'e, Kleyna, \& van der
Marel 2008 for a review).

Points of weakness or controversy regarding Schwarzschild modeling include:
the non-uniqueness of the initial conditions used to generate orbits; the
amount of smoothing or regularization of the solution that is applied and its
impact on the final results; possible over-interpretation of $\chi^2$ plots
and indeterminacy of best solution; how to deal with incomplete positional
sampling; and large computational costs.

\section{Applicability of Schwarzschild's method to Galactic surveys}

We can consider the Galaxy to be composed of two kinds of structure: (1) a
smoothly distributed and old Galaxy in steady-state equilibrium, and (2) a
perturbed, inhomogeneous Galaxy that changes over relatively short timescales
and is not in dynamical equilibrium.  While the classical Galactic
structures of bulge, disk(s), and halo belong to the first category,
shorter-lived structures such as tidal streams, spiral arms, and disk warps,
all fall into the second one.  Only the background, steady-state Galaxy is
susceptible to Schwarzschild modeling. Fortunately, most of the Galaxy's mass
lies in steady-state structures, so a Schwarzschild model should provide a
useful first approximation to the data.  However, even though the
non-equilibrium mass fraction is small, this component is expected to hold
clues to the history of the Galaxy, and means must be found to model it too.

Modeling data from current surveys such as SDSS and RAVE will prepare us for
modeling the vastly superior Gaia data.  Clever arguments such as those in
Smith, Evans, \& An (2009) can only benefit the applicability of
Schwarzschild's technique by narrowing down the range of possible shapes and
geometries of the underlying gravitational potential, and could even shed
light on the optimal choice of initial conditions for orbit integration.

\begin{acknowledgments}

I thank the American Astronomical Society and
the International Astronomical Union  for travel grants.
I acknowledge support from NASA through Hubble Fellowship grant
HF-01216.01-A, awarded by the Space Telescope Science Institute,
which is operated by the AURA, Inc., under NASA contract NAS5-26555.

\end{acknowledgments}

\vfill\eject
 \title[Kinematic Constraints on Galactic
Structure Models] %% give here short title %%
{Connecting Moving Groups to the Bar and Spiral Arms of the Milky Way }

\author[T. Antoja et al.]   %% give here short author list %%
{T. Antoja$^1$,
O. Valenzuela$^2$, F. Figueras$^1$, B. Pichardo$^2$ \& E. Moreno$^2$}%

\affiliation{$^1$DAM and IEEC-UB, ICC-Universitat de Barcelona, Spain, email: tantoja@am.ub.es\\[\affilskip]
$^2$Instituto de Astronomia de la UNAM, Mexico, email: octavio@astroscu.unam.mx}

\maketitle

\begin{abstract}
 We use test-particle orbit integration with a realistic Milky Way (MW)
potential to study the effect of the resonances of the Galactic bar and
spiral arms on the velocity distribution of the Solar Neighbourhood and other
positions of the disk. Our results show that spiral arms create abundant
kinematic substructure and crowd stars into the region of the Hercules moving
group in the velocity plane. Bar resonances can contribute to the origin of
low-angular momentum moving groups like Arcturus.  Particles in the predicted
dark disk of the MW should be affected by the same resonances as stars,
triggering dark-matter moving groups in the disk.  Finally, we evaluate how
this study will be advanced by  upcoming Gaia data.

\end{abstract}

The MW potential and initial conditions (cold, intermediate and hot disks) of
our simulations are described in \cite{antoja09}.  Next figures show: a)
left: the region of Hercules ($V=-40\kms$) is crowded by the the spiral arms
applied to the cold disk, b) middle: the bar with the hot disk creates groups
at low angular momentum ($V=-100\kms$) with long integration time, c) right:
effects of a model with spiral arms and bar. We propose that a good fit
between the observed velocity field -- see \cite{antoja08} -- and the
simulations requires the combined model under IC1, IC2 and IC3, where the
central and low angular momentum moving groups would appear simultaneously. 

\begin{figure}[h!]
\resizebox{4.4cm}{!}{\includegraphics{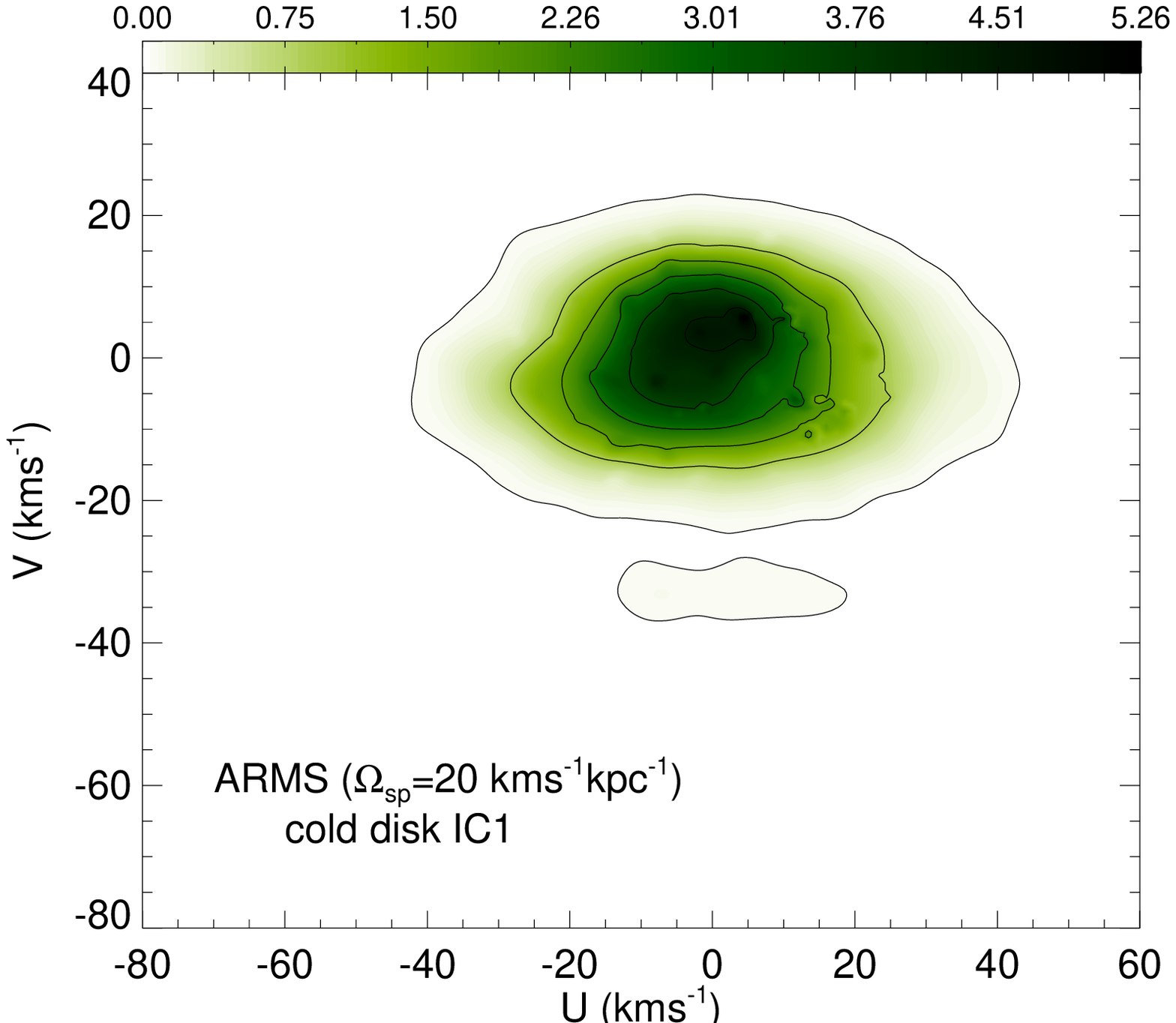}}
\resizebox{4.4cm}{!}{\includegraphics{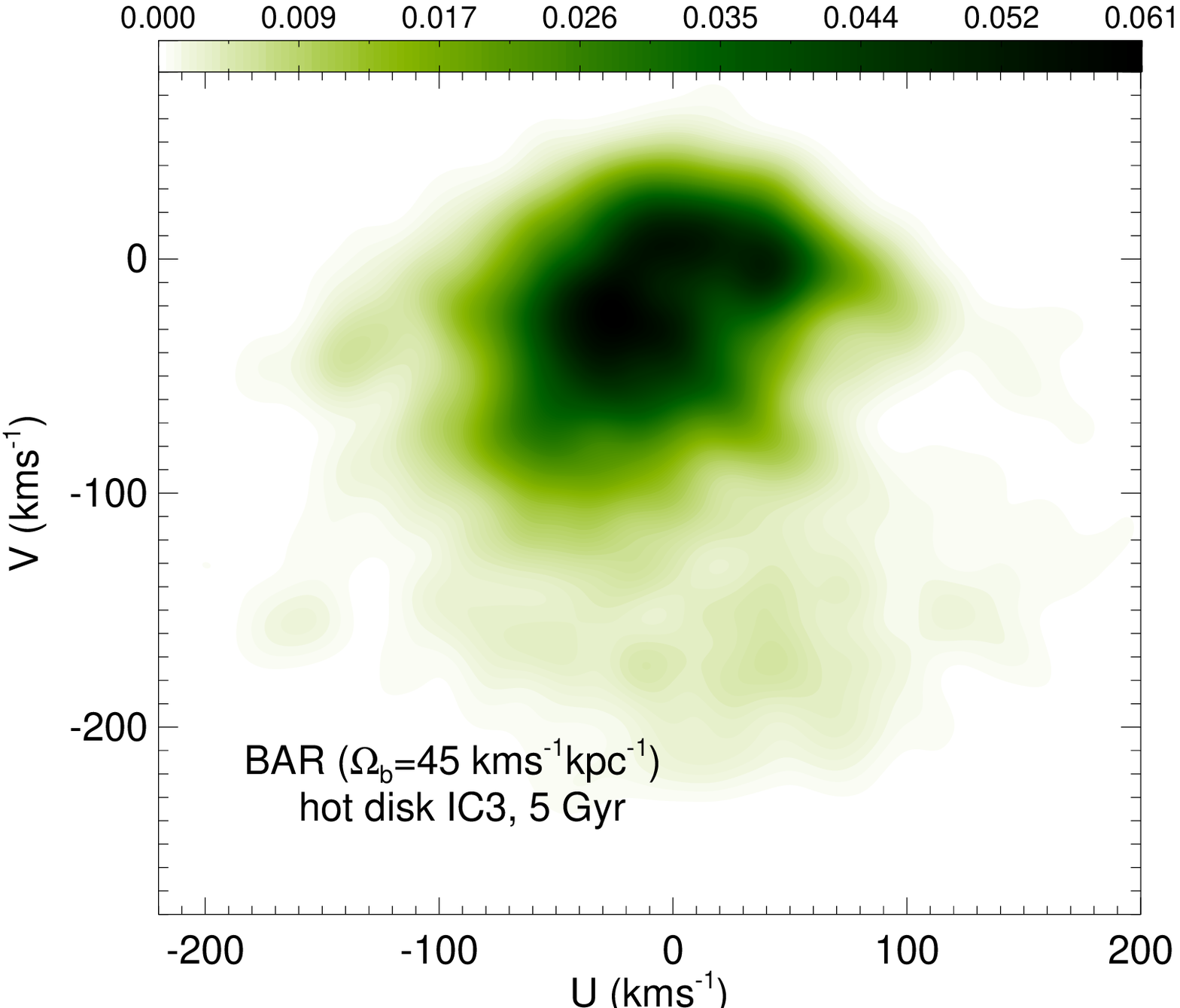}}
\resizebox{4.4cm}{!}{\includegraphics{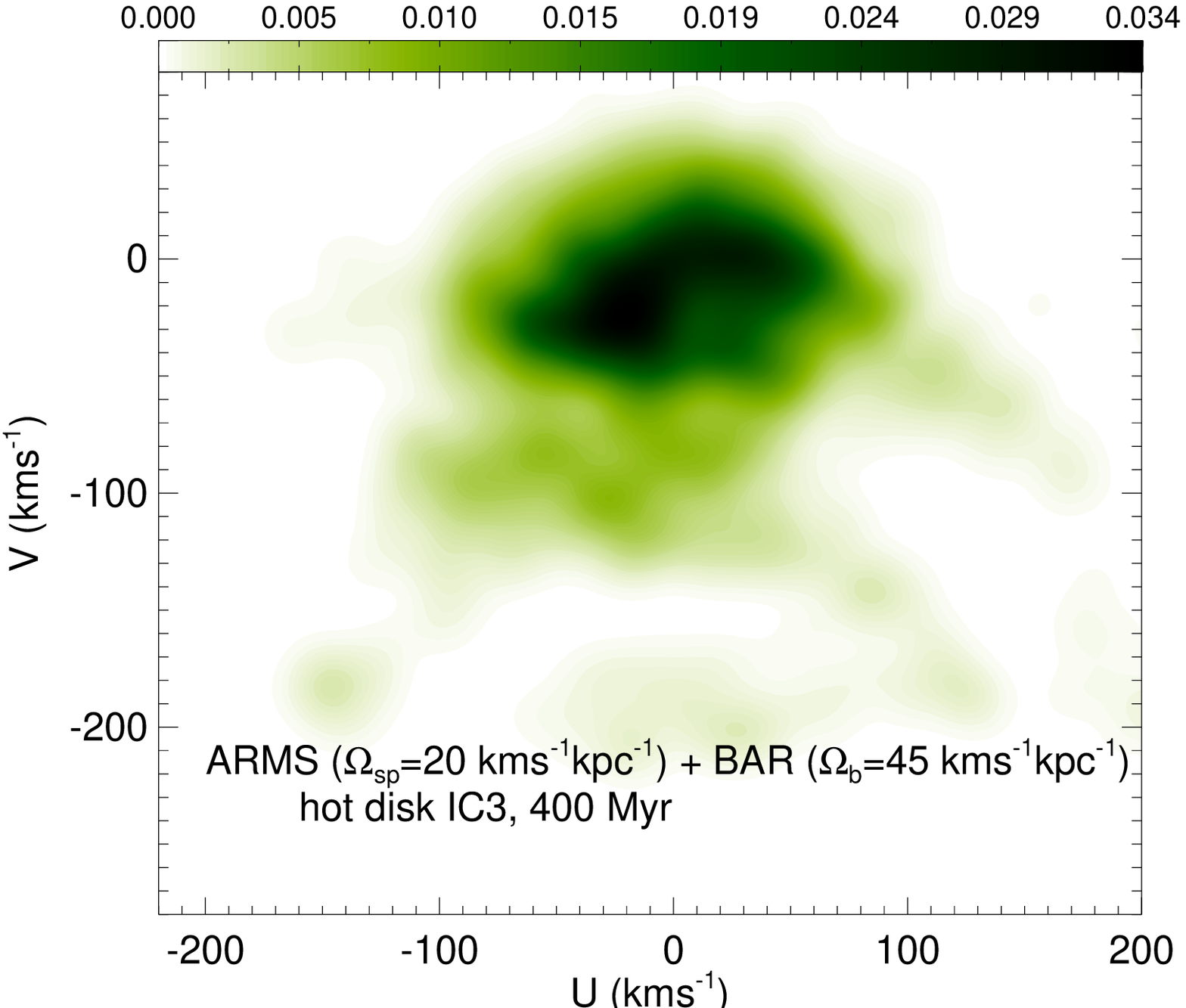}}
\end{figure}

Gaia will revolutionize our knowledge of the Galactic disk. Accuracies in $UVW$
velocities are computed using the Gaia Universe Model Snapshot (GUMS), based
on the Besan\c con Galaxy Model, and the current estimations of the Gaia
errors. We find that we will be able to perform robust statistical analysis
of the velocity distribution (with accuracies better than $2\kms$ in all
components $UVW$) up to $\sim3\,$kpc from the Sun.

\vfill\eject
 \title[Simulations of the Milky Way] %% give here short title %%
{Cosmological simulations of the Milky Way}

\author[Mayer]   %% give here short author list %%
{Lucio Mayer$^1$}

\affiliation{$^1$Institute for Theoretical Physics, University of Zurich, Wintherthurestrasse 190, Zurich (CH)}

\maketitle

\begin{abstract}
Recent simulations of forming low-mass galaxies suggests a strategy for
obtaining realistic models of galaxies like the Milky-Way.
\end{abstract}


Cosmological simulations of galaxy formation are powerful tools for
confronting the $\Lambda$CDM model
with observational datasets.  The increase in mass and spatial resolution
and the improvement of sub-grid algorithms for star formation and feedback
processes have recently resulted in simulated galaxies with realistic disk
size and angular momentum content (Mayer et al. 2008; Governato et al. 2009a).
However, simulated galaxies hosted in halos with masses $\sim 10^{12}
M_\odot$ exhibit prominent bulges and structural parameters reminiscent of
Sa spirals rather than of Sb/Sc galaxies.  Surface densities at the solar
radius are larger than that of the Milky Way (MW) by factors of a few and the
more massive bulge produces a steeper rotation curve compared to that of the
MW (Read et al. 2009).  At  halo masses $M_{\rm
vir} > 2 \times 10^{12} M_{\odot}$ the predominance of hot-mode gas
accretion counters the presence of a prominent star forming disk at $z=0$,
producing earlier-type objects resembling S0 galaxies (Brooks et al. 2009)
and supporting recent estimates based on RAVE that yield $M_{\rm vir} \sim
10^{12} M_{\odot}$ (Smith et al. 2007).

Yet the solution to forming a realistic MW analog could be at hand.  Recently
we have performed galaxy-formation simulations for mass scales $<10^{11}
M_{\odot}$. By sampling these low-mass galaxies with several millions of
particles, we achieve a mass resolution better than $10^3 M_{\odot}$ in the
baryons, thus resolving individual molecular clouds.  Star formation can now
be tied to gas at molecular cloud densities ($\rho > 100$ cm$^{-3}$).  A
realistic, inhomogeneous interstellar medium is obtained that results
naturally in stronger supernova outflows than when the standard
star formation threshold ($\rho = 0.1$ cm$^{-3}$) is adopted.  
Such outflows efficiently remove the
low-angular momentum baryonic material from the central region,
suppressing the formation of a bulge and producing an object with a slowly
rising rotation curve in very close agreement with observed dwarf galaxies
(Governato et al.\ 2009b; see also Ceverino \& Klypin 2009).
We argue that comparable resolution of MW-sized galaxies will
yield rotation curves and bulge-to-disk ratios appropriate for Sb-Sc
spirals at $z=0$.  This  requires  increasing the number of particles
employed by more than an order of
magnitude.

\vfill\eject
 \title[Torus modelling] %% give here short title %%
{Modelling the Galaxy with orbital tori}

\author[J. Binney]   %% give here short author list %%
{James Binney$^1$}

\affiliation{$^1$Rudolf Peierls Centre for Theoretical hysics,
  Keble Road, Oxford OX1 3NP, UK \break email: binney@thphys.ox.ac.uk\\[\affilskip]}

\maketitle

\begin{abstract}
The principles and advantages of torus modelling are explained.

\end{abstract}

\firstsection % if your document starts with a section,
\section{What is torus modelling?}

Torus modelling (\cite[McMillan \& Binney 2008 and references
therein]{McMillanB}) is a modification of Schwarzschild modelling in which
orbits are numerically constructed as three-dimensional surfaces in
six-dimensional phase space rather than as time sequences. These surfaces are
topologically equivalent to 3-tori: if we identify each point of the floor of
a room with the point of the ceiling that is vertically above it, and
similarly identify corresponding points on the front and back walls, and on
the right and left walls, the room becomes a 3-torus. The ``angle variables''
$\theta_1,\theta_2,\theta_3$ are Cartesian coordinates for position in a room
that has been thus made into a 3-torus: for example, as $\theta_3$ increments
from $0$ to $2\pi$, the point moves vertically from floor to ceiling.
Remarkably, as a star orbits through the Galaxy, its angle variables increase
linearly in time, $\theta_i(t)=\theta_i(0)+\Omega_it$, so its point moves in
a straight line through its room-like orbital torus. Unless the frequencies
$\Omega_i$ are rationally related ($\Omega_1:\Omega_2:\Omega_3=n_1:n_2:n_3$
for integer $n_i$) the star eventually comes arbitrarily close to every point
of its torus.

The natural labels for a torus are the three actions
$J_i=(2\pi)^{-1}\oint_{\gamma_i}\vv\cdot\d\vx$, where $\gamma_i$ is the path
on which $\theta_i$ increments from $0$ to $2\pi$ with the other two angle
variables held constant. Indeed the set of six coordinates $(J_i,\theta_i)$
are canonical coordinates for phase space. In particular the Poisson bracket
of any two angle variables vanishes, $[\theta_i,\theta_j]=0$, so tori are
null in the sense that the Poincar\'e invariant of any part of a torus
vanishes ($\int\d\vv\cdot\d\vx=0$).

The idea of torus-modelling is to use software that returns an orbit in
terms of $\vx(\vtheta)$ and $\vv(\vtheta)$ rather than $\vx(t)$ and $\vv(t)$
as a Runge--Kutta integrator does. Since in this picture the orbit is
identified with the torus, it is labelled by the actions $\vJ$.
The following benefits flow from expressing orbits in this way.

\begin{itemize}
\item Given a point in space $\vx$ we can readily find the values of
$\vtheta$ at which the star reaches that point and read off the velocities
$\vv(\vtheta)$ with which the star passes through that point. By contrast, if
we are given $\vx(t)$, we will in general search in vain for a time when the
star is precisely at the given point and will have to settle for times when it
is near. It will not be clear whether we have found times that give
approximations to all the possible values of $\vv$ at the given point.

\item When we integrate orbits in time, the orbit is characterised by its
initial conditions. The same orbit corresponds to infinitely many different
initial conditions, so it is not clear how to do a systematic survey of phase
space to obtain a representative sample of orbits. This difficulty does not
arise in torus modelling because the actions are essentially unique labels of
orbits and the orbits with actions in the range $\d^3\vJ$ occupy a volume
$(2\pi)^3\d^3\vJ$ in 6-d phase space.

\item Actions are adiabatic invariants so we can relate orbits in different
potentials: orbits with the same actions will deform into each other if we
slowly deform the potential from that of one model to that of another. No
such identification is possible if orbits are characterised by their initial
conditions. On account of adiabatic invariance, the distribution function
(DF) of a system such as a globular cluster is invariant as, for example, loss
of gas modifies the cluster's density distribution and gravitational
potential. Together with the previous item, adiabatic invariance make it
straightforward to specify a galaxy model uniquely and to compare the orbital
structures of models that have slightly differing potentials.

\item Analytic formulae for $\vx(\vtheta)$ and
$\vv(\vtheta)$ can be specified using much less data than are required to
specify $\vx(t)$ and $\vv(t)$. Consequently tori greatly simplify
manipulation of orbit libraries. Moreover, expressions for infinitely many
tori can be obtained by interpolating between data for numerically obtained
tori.

\item There is a simple, intuitive connection between the functional form
$f(\vJ)$ of the DF and the real-space properties of the system it specifies
(\cite[Binney \& Tremaine 2008 \S4.6]{BT}).  Moreover, given $f(\vJ)$ there
is a stable scheme for evaluating the self-consistent gravitational
potential, which is not always the case when a DF of the form $f(E,\ldots)$
is specified.

\item The orbit-averaged Fokker-Planck equation takes an exceptionally simple
form in action space (\cite[Binney \& Tremaine 2008 \S7.4.2]{BT}), which should facilitate modelling
of secular evolution.

\item A set of orbital tori specify an integrable Hamiltonian which is very
close to the true Hamiltonian. Perturbation theory works wonderfully well when
this Hamiltonian is used as the point of reference (\cite[Kaasalainen
1994]{Kaasalainen94}).
\end{itemize}

\section{How do we obtain tori?}

Analytic potentials (principally the potentials of the isochrone sphere and the
multi-dimensional harmonic oscillator) provide analytic tori $\vx(\vtheta)$,
etc. We choose such a potential and refer to its structures as ``toy'' ones. A
canonical transformation is used to map the toy torus with given actions
$\vJ'$  into the target phase space. The image torus is guaranteed to be null
but in general it will not lie within a hypersurface on which the target
Hamiltonian is constant, as an orbital torus must. We numerically adjust the
coefficients that define the canonical transformation so as to minimise the
rms variation $\Delta H$ in the target Hamiltonian over the image torus.
Once $\Delta H$ is small enough, the target torus provides an excellent
approximation to an orbital torus.

The canonical transformation $(\vJ,\vtheta)\rightarrow(\vJ',\vtheta')$ is
specified by its generating function
$$
S(\vJ',\vtheta)=\vJ'\cdot\vtheta+\sum_\vn S_\vn(\vJ')\e^{\i\vn\cdot\vtheta},
$$
 where the sum is in principle over all vectors with integer components. The
first term on the right generates the identity transformation, and the
machine has to choose the coefficients $S_\vn(\vJ')$ which are characteristic
of the given orbit. Typically, a good approximation to an orbital torus can
be obtained with  a few tens of non-zero $S_\vn$.

\vfill\eject
 \title[The Milky Way in $N$-Body Models] %% give here short title %% 
{In What Detail Can We Represent the Milky Way in a Conventional $N$-Body Model?}
\author[Debattista]   %% give here short author list %%
{Victor P. Debattista$^1$%
\affiliation{$^1$Jeremiah Horrocks Institute, University of Central
Lancashire,  
Preston, PR1 2HE, UK \break email: vpdebattista@uclan.ac.uk\\[\affilskip]
}}

\maketitle

\begin{abstract}
After a brief review of past $N$-body models of the Milky Way, I
consider some of the difficulties that are inherent in the $N$-body
approach to modelling any disk galaxy.
%\keywords{}
\end{abstract}

\firstsection % if your document starts with a section,
\section{Past $N$-Body Models of the Milky Way}

Modern efforts at modelling the Milky Way (MW) take into account that it is
barred.  The first large effort in this regard was carried out by
\cite{fux97}, who ran simulations of initially axisymmetric
disk$+$bulge$+$halo systems.  He then compared regularly spaced outputs with
a de-reddened {\it COBE} $K$-band bulge map.  He scaled the kinematics of his
model by requiring that the velocity dispersion matches that of Baade's
window.  A number of models provided reasonable fits to the data.  His best
fit bar angle to the sun-center line, $\psi_{\rm bar} = 28^\circ \pm
7^\circ$. Later he added SPH gas to his simulations and was able to fit a
number of features in the CO gas $l-V$ diagram \cite{fux99}.  He argued that
the high-velocity connecting arm is due to the shocked gas within the near
side of the bar.  The gas distribution is also sensitive to the pattern speed
of the bar, and he constrained this parameter to $\Omega_b \sim 50\kmskpc$.

\cite{widrowetal08} modelled the Milky Way by matching observations
(including the rotation curve, local force field, Oort's constants,
local and bulge velocity dispersions, surface density and total mass
within 100$\,$kpc) to a suite of {\it axisymmetric} models.  $N$-body
models of these then all produced bars.  The time of bar formation
depended on the Toomre $Q$ and $X$ parameters; in all cases $\Omega_{\rm p}$
started declining after the bar formed and a dynamically young bar is
required if $\Omega_{\rm p} = 50\kmskpc$.  A problem with this approach
is, however, that bar formation leads to the model departing from
the observations.

$N$-body models are also very useful for testing models constructing
using other methods.  \cite{zhao_96} tested his Schwarzschild model of
the MW bulge using $N$-body simulations, finding that its shape and
mass distribution is stable.

\section{Fundamental Limitations}

Modelling the MW, or any other disk galaxy, by $N$-body simulations is
complicated by a number effects.  Foremost, disk simulations in which
a bar forms are subject to considerable stochasticity.  \cite{sd09}
show that disk simulations which differ only in the seed of the random
number generator used to set up the disk particles evolve quite
differently.  They identified a number of sources of stochasticity,
including multiple disk modes, swing-amplified noise, variations in
the onset and strength of bending instabilities, metastability due to
upward fluctuations in $\Omega_{\rm p}$ \cite[(Sellwood \& Debattista 2006)]{sd06}, and intrinsic chaos.
Stochasticity is weaker when the halo is very massive, but is never
absent.  Such stochasticity makes it hard to improve $N$-body models
by iterating runs with varying parameters.

Modeling is also complicated by radial migration of stars caused by
transient spirals \cite[(Sellwood \& Binney 2002)]{sb02}.  \cite{roskar08b} show that this
migration leads to significant mixing of stellar populations so that
the age distribution of stars at any given radius does not reflect the
star formation history at that radius.  In the solar neighborhood,
\cite{roskar08b} estimate that as much as half the stars formed
elsewhere.  Since the incidence of spirals is chaotic, matching the
stellar populations in simulations requires a certain degree of luck.

\begin{figure}
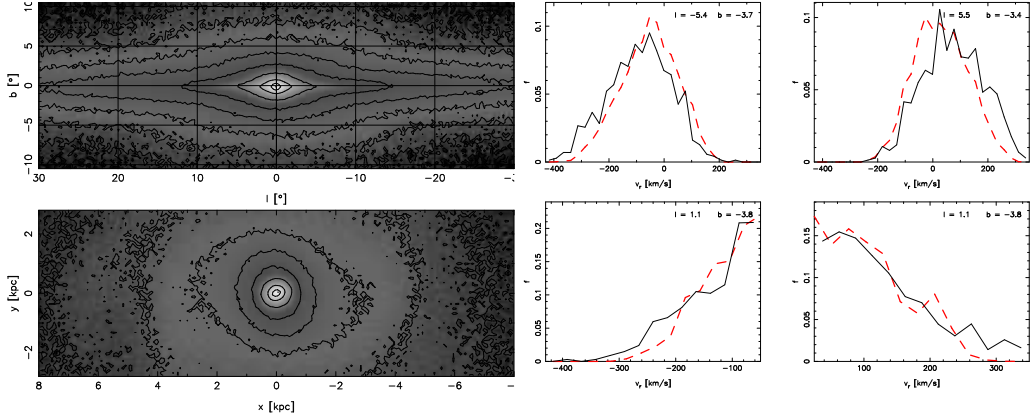

\centerline{
\includegraphics[angle=-90.,width=0.5\hsize]{debattistaf1a.ps}
\includegraphics[angle=-90.,width=0.5\hsize]{debattistaf1b.ps}
}
\caption{$N$-body model (left hand panels) showing the rather poor fit
to the MW density.  Right panels: kinematics for this model (solid
lines) compared with observations (dashed lines).}
\label{fig:MW_kine}
\end{figure}

A third difficulty with modeling the MW is the somewhat weak
constraints that kinematics of the bulge region impose on models.  In
order to demonstrate this, in Figure \ref{fig:MW_kine} I present an
arbitrary disk-galaxy simulation, scaling its velocities to produce a
rotation velocity of 220 km/s.  The density distribution is a rather
poor match to models of the MW [e.g. \cite{bissgerh02,
lop-cor_etal_05}].  However, comparing the kinematics of particles
selected to lie in the bulge using selection functions that match
those in observations of \cite{ran_etal_09} results in distributions
of stellar velocities that are not substantially different from those
observed.

\begin{acknowledgments}
I would like to thank the organizers for inviting me to give this
review.  The author acknowledges support through travel grant
2009/R1 from The Royal Society.
\end{acknowledgments}

\vfill\eject
 \title[Made-to-Measure Modeling of the Milky Way] %% give here short title %%
{Made-to-Measure N-body Modeling of the Milky Way Galaxy}

\author[Ortwin Gerhard]   %% give here short author list %%
{Ortwin Gerhard$^1$}
%   \thanks{Present address: Fluid Mech Inc., 24 The Street, Lagos, Nigeria.},

\affiliation{$^1$Max-Planck-Institut f\"ur Ex.~Physik, Giessenbachstrasse 1,
D-85748 Garching, Germany \break email: gerhard@mpe.mpg.de}

\maketitle

\begin{abstract}
  In this talk a brief introduction is given to made-to-measure
  particle methods and their potential use for modeling the Milky Way
  Galaxy.
%\keywords{Keyword1, keyword2, keyword3, etc.}
\end{abstract}

\firstsection
\section{Made-to-Measure N-body models}

The term ``Made-to-Measure'' (M2M) for a dynamical model reproducing a
set of observational data for a galaxy was coined by
\cite{SyerTremaine96} (hereafter ST). M2M models can be based on
distribution functions, moments, orbits, or particles, but in
particular ST described an algorithm for constructing N-body
equilibrium models, based on the idea of adjusting the masses or
weights of the particles until the system agreed with a prescribed
density distribution.  The central part of their algorithm is the
``force-of-change'' (FOC) equation through which the particle weights
are adjusted according to the mismatch of model and target density
observables. Other important ingredients include a time smoothing to
reduce particle noise and an entropy term to reduce large fluctuations
in the weights.

As a first practical application of this method \cite{Bissantz+04}
constructed a dynamical model for the Milky Way's barred bulge and
disk, constraining the projected density map. However, the ST
algorithm is not well-suited for mixed density and kinematic
observables, and it does not allow a proper treatment of observational
errors. A modified $\chi^2$M2M algorithm which resolved both these
problems was introduced by \cite{DeLorenzi+07} (DL+07) who also
demonstrated the potential of the method by constructing particle
models for spherical, axisymmetric, triaxial and rotating target
systems.  Their implementation NMAGIC has since been used by
De Lorenzi et al.\ (2008,2009) (DL+08, DL+09) to construct dynamical
models of elliptical galaxies based on photometry, Sauron, slit and
planetary-nebula velocity data, and thus to explore the distribution
of dark matter in these galaxies. For modeling the discrete
velocities, \cite{DeLorenzi+08} introduced a likelihood scheme into
the FOC equation, and they also added a separate equation for
adjusting the mass-to-light ratio of the stellar system simultaneously
with the observables.

A modified M2M method was introduced by \cite{Dehnen09}. Rather than
time-averaging the moments of the N-body system as ST and DL+07, he
considered time-averaging the merit function, resulting in a
second-order equation for adjusting the particle weights. This 
contains a linear damping term that acts to maximize the merit
function. Writing the equations in dimensionless time (in
units of orbital period) also made it possible to achieve a uniform
adjustment rate per orbital time for the particle weights, despite
the large range of orbital time-scales inherent in an N-body system. 
\cite{Dehnen09} used this algorithm to construct triaxial equilibrium
target systems.

\section{Comparisons with Schwarzschild and direct 
N-body methods}\label{sec:comp}

M2M particle models include aspects of both Schwarzschild orbit
superposition methods and direct N-body simulations. When the
gravitational potential of the target system is held fixed, the
process of finding a distribution of particle weights to match a
prescribed density field or set of kinematic observables for a galaxy,
is closely related to the process of finding a distribution of orbital
weights in Schwarzschild's method to fit the same constraints.
Conversely, when the adjustment of particle weights is switched off
and the potential is allowed to evolve, M2M particle codes reduce to
N-body methods. E.g., in the work of DL+07 several equilibria are
found in the first limit, while the stability of the final galaxy
models in DL+08 and DL+09 is tested in the second limit. Of course,
the strength of the M2M particle methods is to self-consistently
evolve the particle weights and their potential simultaneously while
approaching the target data (see applications in DL+08). This is
perhaps the greatest advantage over Schwarzschild and N-body
techniques. However, this is most useful when the particles trace the
mass.  Because there is no direct connection between observables and
the global gravitational potential, external dark-matter potentials
have still to be explored one by one. Compared to Schwarzschild and
N-body techniques, M2M modeling is still in an early stage, and
improvements of both the techniques and the efficiency of the method
are needed.

\section{M2M modeling for the Milky Way}\label{sec:MW}

The goal of dynamical models for the Milky Way is to uncover the
fossil record of its assembly history, as described by, for example,
the orbital distribution of stars of different metallicities or other
population parameters.  Because of the enormous detail expected from
future observations, particularly from Gaia, it is likely that 
interesting (sub)structure in the models would be visible in the data.
Thus techniques with a minimum of simplfying assumptions such as
M2M may be well-suited to understand what these data will be
telling us.

M2M work to date has been confined to the inner Galaxy; see
\cite{Bissantz+04} and Debattista et al.\ (in preparation).  These
models have successfully reproduced as diverse observables as
densities, radial velocity histograms and microlensing event
time-scale distribution, but a comprehensive bulge model is still
pending.

Modeling the nearby Galactic disk has not been tried yet. This has the
problem that only a small fraction of the particles from any such
model will be seen in magnitude limited surveys; see e.g.,
\cite{Brown+05}. However, this may actually not be such a large
problem for M2M methods such as NMAGIC.  The time averaging of the
observables allows suppressing the particle noise in the model
observables by a factor 10-100, which eases the comparison with solar
neighbourhood data.

In conclusion, made-to-measure modeling for the Milky Way has little
history, but a lot of promise, and so there is a lot of work to do!

\vfill\eject
 \title[Extinction models of the Galaxy] %% give here short title %%
{How to build a 3D extinction model of the Galaxy}

\author[L\'epine \& Am\^ores]   %% give here short author list %%
{J.R.D. L\'epine$^1$ \& E.B. Am\^ores$^1$}
  
\affiliation{$^1$Instituto de Astronomia, Geofisica e Ci\^encias Atmosf\'ericas, Universidade
 de S\~ao Paulo, Cidade Universit\'aria, S\~ao Paulo, SP, Brazil;
email: jacques@astro.iag.usp.br\\[\affilskip]}

\maketitle

\begin{abstract}

We show that anomalous extinction (deviations from the traditionally  adopted
$R_V = A_V/E(B-V) =3.1$ introduces large uncertainties in the distances of stars, 
for distances larger than 1-2$\,$kpc. We  argue that for such distances and for
directions close to the galactic plane, the use of extinction models based on the 
gas distribution in the Galaxy is safer, for the moment, than the use of extinction maps.

\end{abstract}

\firstsection % if your document starts with a section,
\section{Introduction} It is usually believed that all that we need to derive
the photometric distance of a star, taking into account extinction, is the
color excess $E(B-V)$.  In reality the existence of anomalous extinction
seriously challenges this belief. Besides this, there are two opposite views
concerning the possibility of corrections for interstellar extinction based
on models.  One is that the dust distribution is clumpy and random.
Consequently, it does not make any sense to produce models; only empirical
extinction maps or tables are useful.  The other idea is that the
proportionality between gas and dust column densities is well established.
Simple models of the gas distribution can be constructed, based on HI and CO
surveys, so that extinction is predictable to some extent.  \cite{Amores05}
made an extensive comparison of an extinction model based on the gas
distribution with a large sample of stars with known extinctions (the sample
of \cite[Neckel \& Klare 1980]{Neckel80}). The comparison proved that the
extinction is largely predictable. Interestingly, Neckel \& Klare themselves
produced extinction maps based on their sample of stars, which constitutes an
opportunity for a comparison between the two approaches. 

\vspace{-3 mm}
\section{The problem of anomalous extinction, and discussion}

\cite{Neckel80} in a famous work, determined the spectral type and color
excess for more than 7000 O and B stars situated near the galactic plane.
They plotted the interstellar extinction $A_V$ as a function of distance for
many directions, and obtained some unexpected results. In many cases, one can
observe in those plots a low extinction up to a distance of the order of
$1\,$kpc (see eg. the direction $l = 128,\, b= 0$), followed by a step in
the extinction up to A$_V$= 2-3, and then $A_V$ remains constant up to about
5$\,$kpc, the maximum distance investigated.  It is not surprising to see steps
in $A_V$, as they are explained by the presence of dense clouds along the
line-of-sight.  What is surprising is that there are paths of many kpc
without steps, as if all the clouds were close to the Sun. Another surprising
result, noted by Neckel \& Klare, is that the clouds situated close to the
Sun are bigger than the more distant ones. 

The explanation for the two unexpected results is that the distances of the
more distant stars (and clouds) are overestimated. The clouds 
contain dense cores in which $R_V$ can be much larger than the classical value 3.1.
As a consequence, the extinction is underestimated and the distances are 
overestimated. The existence of anomalous extinction is not a new result, 
we only call attention to the fact that the anomalous extinction is so 
widespread that it affects strongly our understanding of the local structures. 

We present in Figure 1  the photometric distances of the sample of OB stars
of the solar neighborhood taken from the Hipparcos catalog. We first 
computed the distances based on the absolute magnitudes expected from their
spectral class, with the extinction estimated using $R_V$= 3.1. We also computed
the distances based on the infrared H magnitudes taken from 2Mass
and the relation $A_H=0.18 E(V-H)$ derived from the extinction relations by
\cite[Koornneef 1983]{Koornneef83}. The distances using V and H bands are not different, the
ones based on H magnitudes are shown in Figure 1.
Since the extinction in band H is smaller than in V, we would
expect that the anomalous extinction would also be smaller. However, it can be
seen in the figure that elongated structures like fingers of God are 
present. The elongated structures cannot be the result of any kind
of calibration errors. They are the result of an unpredictable scattering in the
values of $R_V$. The distance of the first interstellar cloud found along a 
line-of sight is correct, since the stars are still unaffected by 
extinction, but the next clouds have distances overestimated. Extinction maps,
to be useful for a 3D description of the extinction in the Galaxy, would have to 
give the distances of all the steps in $A_V$, which is not possible at the moment.

How to correct for anomalous extinction? Of course when the data from Gaia
become available we will be able to place real distances to the molecular clouds
which are responsible for steps in the extinction, up to much larger distances.
Meanwhile, it is possible that the approach of \cite{Fitzpatrick09} which describes 
$R_V$ in terms of two color indices could be a good solution.

\begin{figure}
\centering
\includegraphics[height=2in]{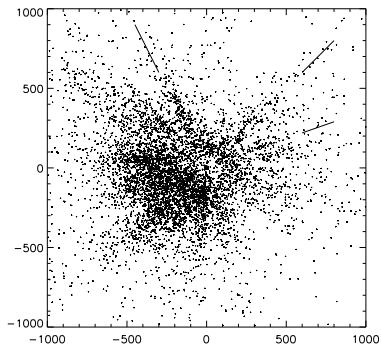}
  \caption{OB stars in the plane of the Galaxy. The distances from the Sun are
   indicated in parsecs, the centre of the Galaxy is outside the figure, at
   $(0, -7500)$. The line segments point to examples of directions of alignments 
   of stars or ``fingers of God" caused by anomalous extinction.
   }
\end{figure} 
\vspace{-2 mm}
\begin{acknowledgments}
We thank  Leticia Vaz and Nayara Torres for the compilation of the catalog of OB stars.
\end{acknowledgments}

\vfill\eject
 \title[A new approach to models of our Galaxy] %% give here short title %%
{A New Approach to the Construction of Dynamical Structure of our Galaxy}

\author[Ueda et al.]   %% give here short author list %%
{H. Ueda$^1$ N. Gouda$^2$  T. Yano$^2$  H. Koyama$^3$ 
  \break \and M. Sakagami$^4$}

\affiliation{$^1$Faculty of Education and Human Studies, Akita University, Tegata-gakuen, Akita 010-0852, Japan \break email: ueda@ipc.akita-u.ac.jp \\
$^2$National Astronomical Observatory, Mitaka, Tokyo 181-8588, Japan \break email:naoteru.gouda@nao.ac.jp, yano.t@nao.ac.jp  \\
$^3$Graduate School of Human and Environmental Studies, Kyoto University, Kyoto 606-8501, Japan \break email: sakagami@grav.mbox.media.kyoto-u.ac.jp \\
$^4$Department of Physics, Waseda University, Shinjuku, Tokyo 169-8555, Japan \break email: koyama@gravity.phys.waseda.ac.jp \\
}

\maketitle

Information about positions and velocities of stars that will be 
gained in the era of GAIA is crucial for determining dynamical 
structure in our Galaxy. The distribution function 
of all component objects in our Galaxy is fundamental for describing 
its dynamics. However, only the distribution 
function of observable stars is obtained from space astrometry 
observations, and it is 
therefore necessary to develop theoretical studies of how to 
construct the distribution function of all matter 
including dark matter and unobservable stars using astrometric
data of observable stars. 
This procedure falls into three categories.
%\\

\noindent
1) Torus Fitting:
\indent
In the first step, we find action variables ${\bf J}$ in any gravitational
potential (Hamiltonian) whose system is almost integrable and considered to
be a representation of our Galaxy.
Concretely,  we find the coordinate transformation 
between action variables and the Cartesian coordinates 
$({\bf x},{\bf p})$ in a model of our Galaxy.
Although the action-angle variables $({\bf J},{\bf \theta})$ provide the most 
compact representation of a regular orbit, it is impossible to
get analytically the coordinate transformation
$({\bf J},{\bf \theta}) \Leftrightarrow ({\bf x},{\bf p})$  
in an arbitrary potential.
We therefore propose a new approach that numerically obtains action 
variables ${\bf J}$ and coordinate transformations
$({\bf J},{\bf \theta}) \Leftrightarrow ({\bf x},{\bf p})$.
It is called a ``Torus Fitting method" and is an alternative
of the Torus Construction method proposed by Binney and his collaborators.
%\\

\noindent 2) Determination of a theoretical distribution function by M2M:
\indent In the second place, we construct the distribution function $f_{\rm
matter}(J_1,J_2,J_3)$ of all matter in the system, using action variables
from the first step. We do this using the made-to-measure algorithm for
constructing an $N$-body realization of an equilibrium stellar system (please
refer to Yano's presentation for details). 
 %\\

\noindent 3) Determination of a model of our Galaxy from the distribution
function of observed stars, $f_{\rm obs}({\bf x},{\bf p})$: \indent Finally,
we have to determine a model of the gravitational potential (Hamiltonian) of
our Galaxy.  Hence we need to construct the distribution function
of all matter including dark matter and unobservable stars using astrometric
data that include the error in distance measurements and
selection biases.  In this procedure a technique that utilizes Hermite
polynomials is used.

From the above procedure we obtain the distribution function 
of all matter using space astrometry observations.  
\vfill\eject
 \title[short title of paper] %% give here short title %%
{An iterative method for constructing stellar systems models: 
how far does it work?}

\author[Sotnikova and Rodionov] %% give here short author list %%
{Natalia Ya. Sotnikova and Sergei A. Rodionov}

\affiliation{Saint-Petersburg State University, Saint-Petersburg,
Russia \break email: nsot@astro.spbu.ru\\[\affilskip]}

\maketitle

\begin{abstract}
We present a new method for constructing equilibrium phase models for stellar 
systems. Applications of the iterative method include both 
modelling of observational data and the construction of initial condition 
for N-body simulations.
\end{abstract}

The aim of the iterative method (IM) is to construct an equilibrium N-body
model with prescribed parameters, or constraints. Setting a given mass
distribution and almost arbitrary velocities of particles, we start the
iterative procedure, by letting the system go through a sequence of
self-consistent evolutionary steps of short duration (iterations).  At the
end of each step, and before the new step is started, we transfer the new
velocity distribution from a bit evolved system to a system with the initial
density distribution. At this stage we need to correct individual particle
velocities in accordance with imposed kinematic constraints (see details in
\cite[Rodionov et al.\ 2009]{RAS09}). We stop iterations when the velocity
distribution ceases to change, which implies that the system has reached
equilibrium.

We managed to construct equilibrium systems of various types -- from spherical 
to triaxial, from one-component to multi-component, from isotropic to 
anisotropic (\cite{RS06}; \cite{SR08}; \cite{RAS09}). 
Successful reconstruction of the distribution function of a model 
disc galaxy from its line-of-sight kinematics encouraged us to use the IM 
to derive the 3D kinematics of edge-on galaxies from observational data. Now 
we have all IR photometric parameters (for a bulge and a disc) of an edge-on 
galaxy NGC~4111 and obtained stellar LSVD of this galaxy at the 6-m telescope. 
Preliminary interpretation of kinematic and photometric observations 
of this galaxy in terms of its 3D structure and 3D velocity distribution 
showed that the IM may be very powerful method to reconstruct  phase-space 
models of real galaxies (Sotnikova et al., 2010, 
in preparation).

\begin{acknowledgments}
This work was supported by the Russian Foundation for Basic Research 
(grant 09-02-00968) and by a grant from President of the RF for support 
of Leading Scientific Schools (grant NSh-1318.2008.02).
\end{acknowledgments}

\vfill\eject
\end{document}